\documentclass[runningheads]{llncs}
\usepackage[colorlinks=true,
            linkcolor=red,
            urlcolor=blue,
            citecolor=blue]{hyperref}
\usepackage{graphicx}
\usepackage{amssymb}
\usepackage{wrapfig}
\usepackage{graphicx}
\usepackage{balance}
\usepackage{color}

\usepackage{xcolor}
\usepackage{multirow}
\usepackage{subcaption}

\let\oldmaketitle\maketitle
\renewcommand{\maketitle}{\oldmaketitle\setcounter{footnote}{0}}

\begin{document}
\title{HHAR-net: \underline{H}ierarchical \underline{H}uman \underline{A}ctivity \underline{R}ecognition using Neural \underline{Net}works}

\titlerunning{HHAR-net: Hierarchical Human Activity Recognition}
%








\author{Mehrdad Fazli\inst{1} \and
Kamran Kowsari\inst{1,2} \and
Erfaneh Gharavi\inst{1} \and\\
Laura Barnes\inst{1} \and
Afsaneh Doryab\inst{1}}

\authorrunning{M. Fazli, K. Kowsari et al.}
%
\institute{University of Virginia, Charlottesville VA 22903, USA 
\and
University of California, Los Angeles CA 90095, USA\\\email{\{mf4yc, kk7nc, eg8qe, lb3dp, ad4ks\}@virginia.edu}
}
\maketitle              

\begin{abstract}
Activity recognition using built-in sensors in smart and wearable devices provides great opportunities to understand and detect human behavior in the wild and gives a more holistic view of individuals' health and well being. Numerous computational methods have been applied to sensor streams to recognize different daily activities. However, most methods are unable to capture different layers of activities concealed in human behavior. Also, the performance of the models starts to decrease with increasing the number of activities. This research aims at building a hierarchical classification with Neural Networks to recognize human activities based on different levels of abstraction. We evaluate our model on the Extrasensory dataset; a dataset collected in the wild and containing data from smartphones and smartwatches. We use a two-level hierarchy with a total of six mutually exclusive labels namely, “lying down”, “sitting”, “standing in place”, “walking”, “running”, and “bicycling” divided into “stationary” and “non-stationary”. The results show that our model can recognize low-level activities (stationary/non-stationary) with 95.8\% accuracy and overall accuracy of 92.8\% over six labels. This is 3\% above our best performing baseline\footnote{HHAR-net is shared as an open source tool at \url{https://github.com/mehrdadfazli/HHAR-Net}}.

\end{abstract}
\section{Introduction}
Human Activity Recognition (HAR) applied to data streams collected from mobile and embedded sensors~\cite{Lara2013,Bulling2014,Chen2012} has numerous real-world applications in understanding human behavior in the wild for health monitoring~\cite{afsaneh2011} (mental and physical), smart environments~\cite{doryab2012activity,bardram2011phase}, elderly care~\cite{hong2008activity,jalal2014depth} and sports applications~\cite{inproceedings,direk2012team}.

The advancements in mobile and wearable devices have made it possible to collect streams of data from built-in sensors in smartphones and fitness trackers including accelerometer, gyroscope, magnetometer, GPS, and microphone. These sensor streams have been analyzed and modeled individually or in combination to recognize basic human activities such as running, walking, sitting, climbing stairs as well as daily activities such as cooking, shopping, and watching TV ~\cite{Su2014,Bayat2014,5482729,7439743,Wang2019}. Detecting human activities and discovering behavioral patterns sets the ground for real-time monitoring of individuals’ physical and mental health. This has urged researchers in the field of connected health to strive for building more accurate HAR systems.

Different computational methods ranging from classic machine learning (e.g., decision trees, naive Bayes\cite{maurer2006activity,jatoba2008context}, and Logistic Regression~\cite{Vaizman2017}) to graphical models such as HMM~\cite{nguyen2005learning,duong2005activity} and Conditional Random Fields~\cite{nazerfard2010conditional,vail2007conditional} to Neural Networks~\cite{Zeng2014,murad2017deep,Ronao2016} and Pattern Mining~\cite{gu2010pattern,chikhaoui2011frequent,gu2009mining} have been applied to recognize human activities. These methods, however, are unable to recognize different levels of activities and their abstractions in one model. 

In this paper, we present a  tree-based hierarchical classifier using Neural Networks to both improve accuracy and capture  different levels of abstraction. We use the Extrasensory dataset \cite{Vaizman2017} to evaluate our proposed model. We test our approach on 6 activity labels, “Lying down”, “sitting” and “standing” which are stationary activities and “walking”, “running” and “bicycling” which are non-stationary activities. The results obtained on these basic activities are promising and indicate the capability of our method to be applied to more complex HAR systems.

\section{Methodology}
\subsection{Data Processing}
To clean the data, we discard all the samples other than those that have at least one relevant label from the six target labels. We do not remove samples with one or more missing labels for those six target labels as long as one of the six labels is relevant. For example, if the user was sitting at a certain time, it means no other activities could be performed. We also apply mean imputation for the missing features to avoid losing data.

\subsection{Hierarchical Classification}
Hierarchical Classification algorithms employ stacks of machine learning architectures to provide specialized understanding at each level of the data hierarchy~\cite{kowsari2017hdltex} which have been used in many domains such as text and document classification, medical image classification~\cite{kowsari2020hmic}, web content~\cite{dumais2000hierarchical} and sensor data~\cite{mccall2012macro}. Human activity recognition is a multi-class (and in some cases multi-label) classification task. Examples of activities include sitting, walking, eating, watching TV, and bathing. 
The following contrasts the two types of classifiers we use in this paper for multi-class classification, namely \textit{flat} and \textit{hierarchical}. Given n classes as \(C=\{c_1, c_2, ..., c_n\}\) in which \(C\) is the set of all classes, a flat classifier directly outputs one of these classes, whereas a hierarchical classifier first segments the output in \(\{c_p, c_p+1, ..., c_q\}\) and then provides the final label in the next level \(c_i\) where \(p\leq i\leq q\). C. N. Silla Jr., A. A. Freitas~\cite{Silla2011} conducted a comprehensive study on the hierarchical classifications across different fields whose terminology we use in this paper.

Our proposed method is to manually apply the inherent structure that exists in the context (activity) labels to our classifier. For example, activities done by individuals can very well be categorized as stationary and non-stationary activities. Lying down, standing, sleeping, and sitting are all considered stationary activities during which your net body displacement is negligible. On the other hand, activities such as running, walking, swimming, driving, bicycling and working out, are all examples of non-stationary activities where your net body displacement is non-negligible.

Activities can also be divided into indoor and outdoor activities. Indoor activities can be washing dishes, doing laundry and watching TV as opposed to biking and driving that are examples of outdoor activities. However, this classification can be more challenging as many activities can occur both indoor and outdoor. Additionally, many exclusively indoor activities are rare activities that Extrasensory dataset does not have sufficient data for training a model. Hence, we only focused on stationary and non-stationary activities in this paper.

In this experiment, we use the stationary vs non-stationary grouping of six mutually exclusive activities including “lying down”, “sitting”, “standing still”, “walking”, “running”, and “bicycling”. The first three activities are stationary and the rest are non-stationary activities. We also make sure that there is no sample in the dataset that indicates two of them happening at the same time (possibly due to misreporting).

\begin{figure}[tb]
    \centering
    \includegraphics[width=.9\columnwidth]{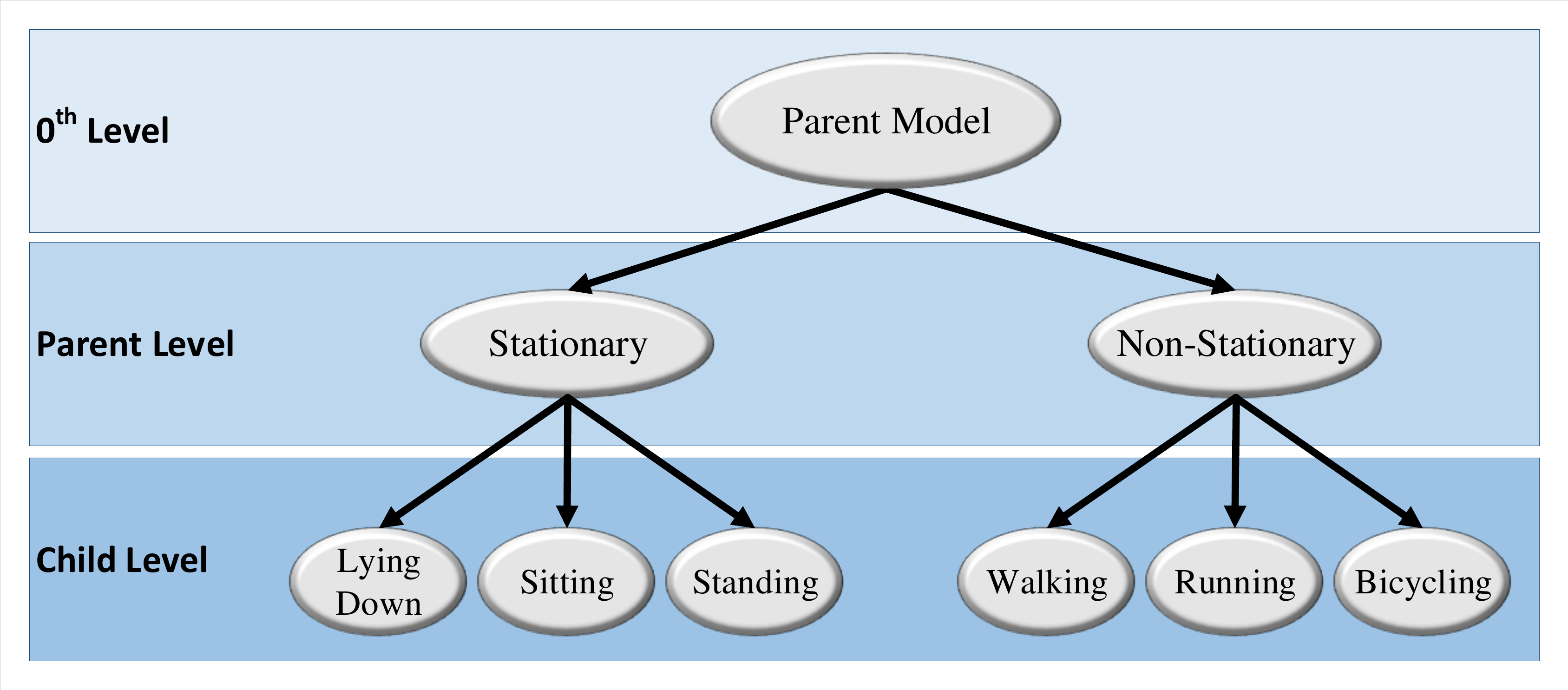}
    \caption{Schematic of the hierarchical model employed for the classification task}
    \label{fig:hierarchical_structure}
\end{figure}

Before introducing a hierarchy into the labels, we train a flat classifier and observe the performance of the classifier carefully. If there is a considerable misclassification between two or more distinct sets of classes, a hierarchy might boost the performance of the model.

As shown in Figure \ref{fig:hierarchical_structure}, we employ a hierarchical classification with one local classifier per parent node (LCPN) \cite{Silla2011}. In this setting, a classifier first classifies activity samples into stationary and non-stationary. In the second (child) level, one algorithm further classifies the stationary samples into sitting, standing, and lying down, and another classifier labels non-stationary activities into walking, running, and biking. This is a two-level hierarchy. The training of the parent nodes are done independently but, in the test phase we use a top-down prediction method to assess the overall performance of the model.

\subsection{Deep Neural Networks}
\begin{figure*}[!b]
    \centering
    \includegraphics[width=\textwidth]{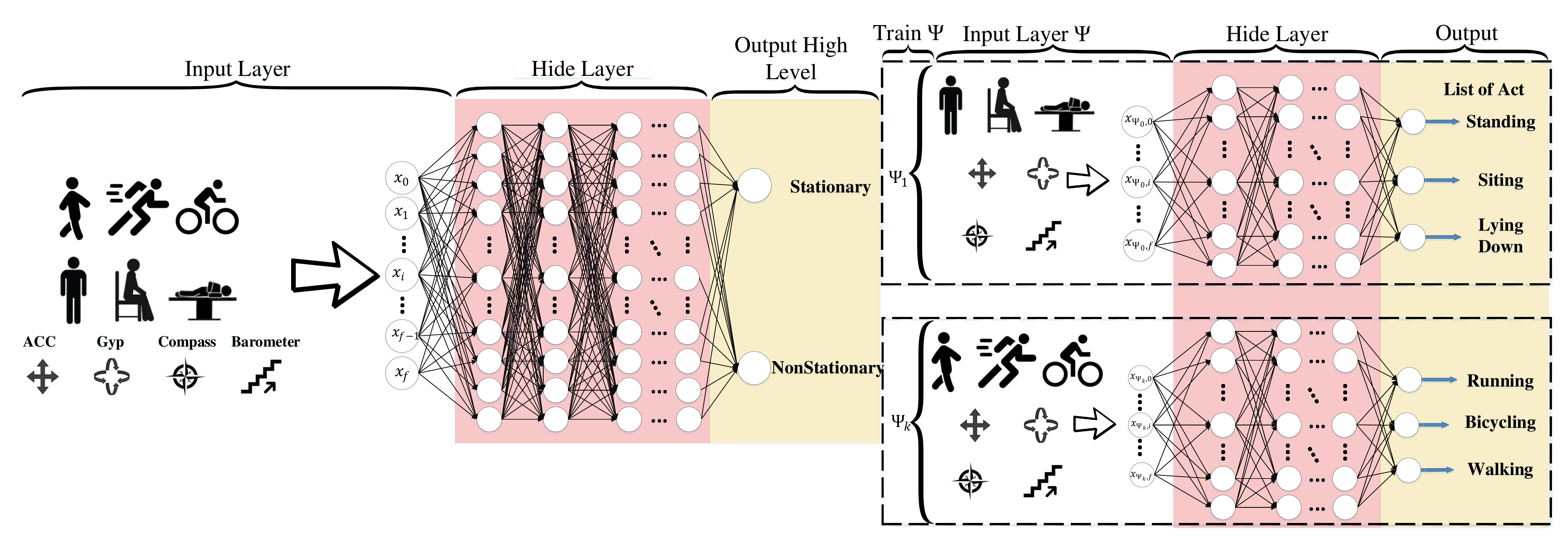}
    \centering
    \caption{HHAR-Net: \underline{H}ierarchical \underline{H}uman \underline{A}ctivity \underline{R}ecognition using Neural \underline{Net}works. This is our Deep Neural Network~(DNN) approach for Activity Recognition. The left figure depicts the parent-level of our model, and the right figure depicts child-level models defined by $\Psi_i$ as input activity in the parent level~\cite{kowsari2017hdltex}.}
    \label{fig:HHAR_net}
\end{figure*}

Deep neural networks (DNN) are artificial neural networks~(ANN) with more than one hidden layer. The schematic architecture of a DNN is shown in Figure~\ref{fig:HHAR_net}. The number of nodes at the input layer is equal to the number of the features and the number of the nodes at the output layer is equal to the number of the classes~\cite{kowsari2018rmdl}. DNNs are capable of finding a complex nonlinear relationship between the inputs and outputs. Also, they have shown tremendous power in prediction if designed and tuned well. Tuning of a DNN consist of tuning the number of the layers, number of the nodes per layer, appropriate activation function, regularization factor and other hyperparameters.

\subsection{Evaluation}
Training and testing of the classifiers in this setup are done independently which calls for creating separate training and test sets for each classifier. Since stationary and non-stationary are not among the labels, we create them using a simple logical OR over the child labels and build a binary classifier to label them. The second and third classifiers are corresponding to stationary and non-stationary nodes at the parent level and both of them are three-class classifiers.

The test data is passed through the first classifier at \(0^{th}\) level and based upon the predicted class it is passed to one of the classifiers at the parent level to make the final prediction out of the six labels. By comparing the actual labels with the predicted ones, we calculate the confusion matrix. Other performance metrics, such as accuracy, precision, recall, and F1 score can be readily obtained from the confusion matrix.

Accuracy score definition can be extended to multi-class classification and can be reported per class or averaged over all classes. 
However, as data becomes more imbalanced, the accuracy can be misleading. Therefore, the balanced accuracy becomes more relevant.

\subsection{Baseline Methods}
To better evaluate our model we needed a baseline to compare our results with. To our knowledge, there is no similar study with the same setup (same labels and dataset). Therefore we applied several classification algorithms along with a flat DNN on our data. Decision tree with a max depth of 20, k Nearest Neighbors (kNN) with k=10, support vector machine (SVM) with "RBF" kernel, random forest with max depth of 10 and 10 estimators, and multi-layer perceptrons (one hidden layer with 64 nodes) were applied on the six class classification problem. The accuracy of these algorithms are compared with our model in the results section.

\section{Experiment}

\subsection{Dataset}
The Extrasensory dataset that is used in this work is a publicly available dataset collected by Vaizman et al. \cite{Vaizman2017} at the University of California San Diego. This data was collected in the wild using smartphones and wearable devices. The users were asked to provide the ground truth label for their activities. 
The dataset contains over 300k samples (one sample per minute) and 50 activity and context labels collected from 60 individuals. Each label has its binary column indicating whether that label was relevant or not. Data was collected from accelerometer, gyroscope, magnetometer, watch compass, location, audio, and phone state sensors. Both  featurized and raw data are provided in the dataset. Featurized data has a total of 225 features extracted from six main sensors. We used the featurized data in this work.


\begin{figure}[!t]
  \centering
  \subfloat[Flat DNN]{\includegraphics[width=0.77\textwidth]{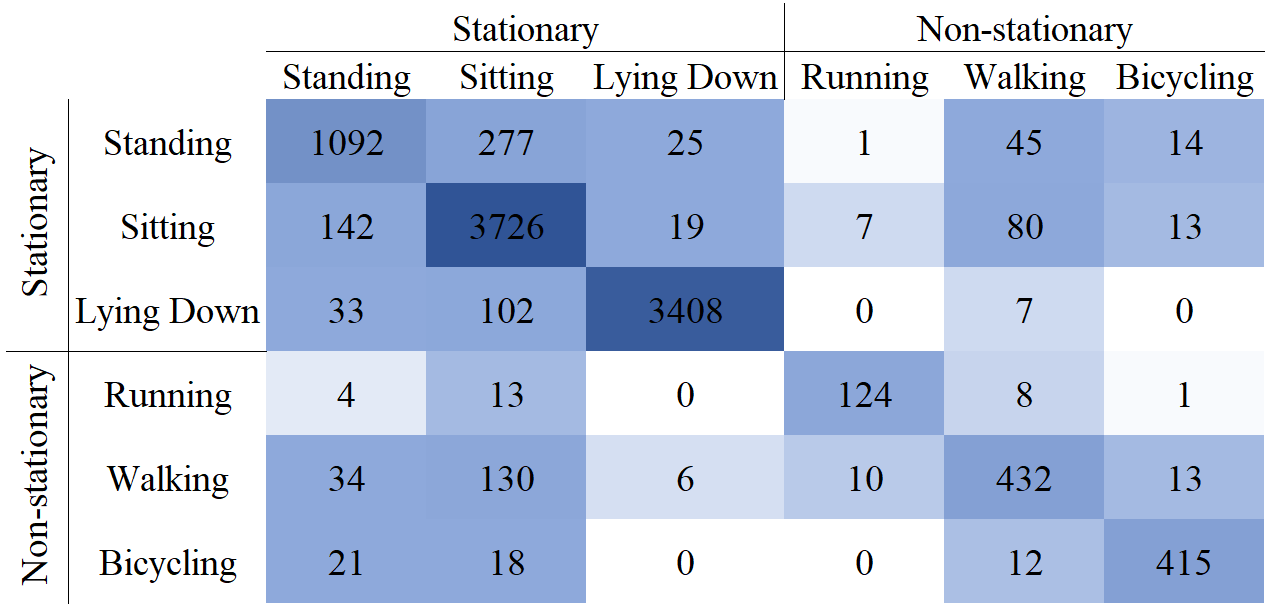}\label{fig:F_Conf_Mat}}
 \vspace{+12pt}
  \subfloat[Hierarchical DNN]{\includegraphics[width=0.77\textwidth]{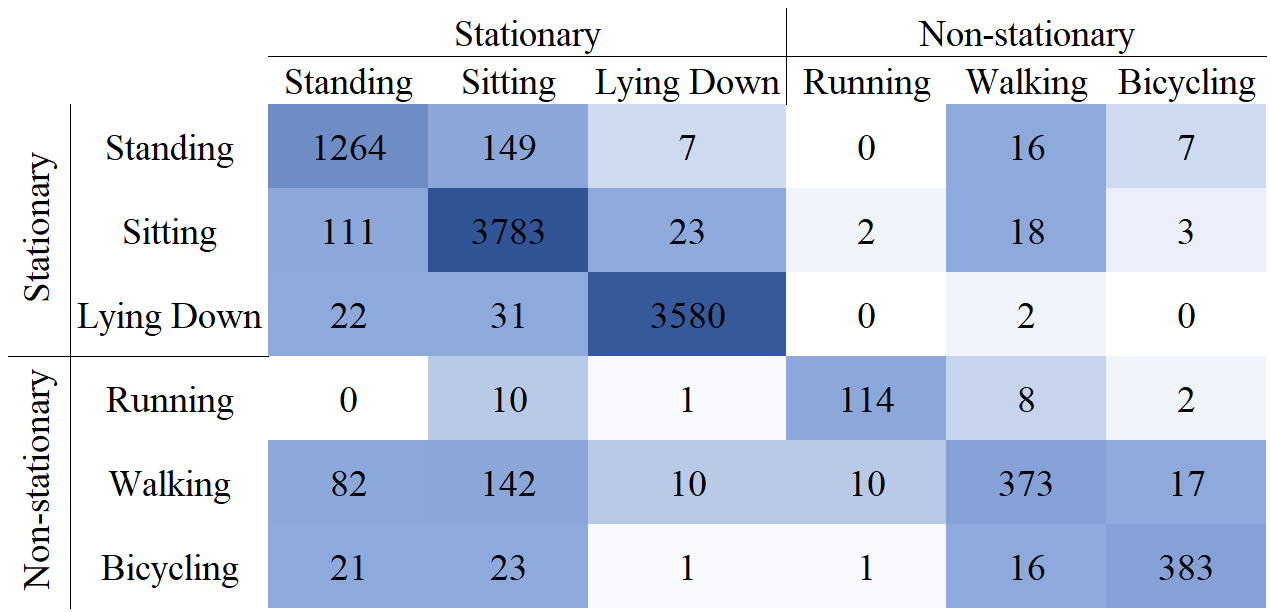}\label{fig:H_Conf_Mat}}
  \caption{Confusion matrix of a) flat classifier and b) hierarchical classifier}
\end{figure}

\subsection{Results and discussion}
As mentioned earlier, we first train a flat classifier and  examine the performance of the classifier before deciding to use the hierarchy. Hence, a DNN with 3 hidden layers and 256, 512, 128 nodes respectively were trained. A dropout layer was also added to prevent overfitting \cite{srivastava2014dropout}. The number of nodes and the number of hidden layers varied between \{2, 3\} and \{64, 128, 256, 512\} respectively. The activation functions for the hidden layers were set to rectified linear unit (ReLU) \cite{nair2010rectified}. In the output layer, softmax was used as the activation function for multi-class classifiers and sigmoid for the \(0^th\) level of the hierarchical model. Class weights were set to the inverse of the frequency of classes to compensate for the class imbalance. The implementation was done in Python using Keras \cite{chollet2015keras} and sklearn \cite{scikit-learn} packages. From the entire dataset, 80\% was selected for the training of the model and 20\% was reserved for the evaluation of the performance of the model. Also, 5\% of the training data was used as an evaluation set to tune hyper-parameters.

The resulting confusion matrix is shown in Figure \ref{fig:F_Conf_Mat}. A quick glance at the confusion matrix reveals that the number of misclassifications between stationary activities and non-stationary activities is not negligible. But, to make a quantitative comparison, we calculated the number of misclassifications, between stationary activities and non-stationary activities and within each of them. To do so, we summarized misclassifications for our flat DNN in Table \ref{table:Flat_Misclassifications}. The absolute number of misclassifications within the stationary class is 598 which is 57.8\% of all misclassifications. It is not a surprise to our intuition as discerning stationary activities are expected to be a challenge. However, the main takeaway from this table, is that nearly 38\% of the total misclassifications are coming from the confusion between stationary and non-stationary classes. This implies that our classifier is struggling with distinguishing between these two types of activities in some cases. Surprisingly, misclassifications between "sitting" and "walking" is contributing to more than 20\% of the total misclassifications. This urged us to come up with a hierarchy to reduce the misclassifications between the two sets of classes.

\begin{table}[!t]
\centering
\begin{subtable}{0.4\textwidth}
\centering
\caption{Flat DNN. Total misclassifications are 1035.}
\begin{tabular}{c c c}
      & ~~~~S~~~~ & ~~~Non-S~~~  \\
     \hline
     S & 598 & 167 \\
     \hline
     Non-S & 226 & 44 \\\hline
     
     \label{table:Flat_Misclassifications}
     \end{tabular}
\end{subtable}
\begin{subtable}{0.4\textwidth}
\centering
\caption{Hierarchical DNN. Total misclassifications are 735}
\begin{tabular}{c c c}
     & ~~~~S~~~~ & ~~~Non-S~~~ \\
     \hline
     S & 343 & 48 \\
     \hline
     Non-S & 290 & 54 \\\hline
     \label{table:Hierarchical_Misclassification}
     \end{tabular}
\end{subtable}%
\caption{Miss-classification within and between stationary (S) and non-stationary (Non-S) classes.}
\end{table}

\begin{table}[!b]
\centering
\begin{subtable}{0.4\textwidth}
\centering
\caption{Accuracy.}\label{table:Accuracy_Benchmark}
\begin{tabular}{ l c c}
\hline
     Classifier & ~~~Accuracy~~~ \\ 
     \hline
     Decision Tree & 84.3 \\ \hline
     k-NN & 87.5 \\ \hline
     SVM & 87.7 \\ \hline
     Random Forest & 83.5 \\ \hline
     MLP & 87.8 \\ \hline
     Flat DNN & 89.8 \\ \hline
     \textbf{HHAR-Net} & \textbf{92.8} \\\hline

     \end{tabular}
\end{subtable}
\begin{subtable}{0.4\textwidth}
\centering
\caption{Balanced Accuracy}\label{table:Balanced_Accuracy_Benchmark}
\begin{tabular}{ l c c}
\hline
     Classifier & Balanced Accuracy \\
     \hline
     Decision Tree & 75.9 \\ \hline
     k-NN & 78.8 \\ \hline
     SVM & 79.2 \\ \hline
     Random Forest & 70.9 \\ \hline
     MLP & 81.4 \\ \hline
     Flat DNN & 84.1 \\ \hline
     \textbf{HHAR-Net} & \textbf{85.2} \\\hline

     \end{tabular}
\end{subtable}%
\vspace{12pt}
\caption{Performance comparison for several flat classifiers and the hierarchical classifier.}
\end{table}

\begin{table}[!t]
\centering

\begin{tabular}{|c|c|c|c|c|c|c|}
\hline
                                  &                                 &            & Precision & Sensitivity & Specificity & F1-score \\ \hline
\multirow{6}{*}{\rotatebox[origin=c]{90}{Flat DNN}}         & \multirow{3}{*}{Stationary}     & Standing   &     82.35$\pm$ 1.95      &    75.10 $\pm$ 2.22        &      97.19$\pm$ 0.84       &      78.56 $\pm$ 2.10    \\ \cline{3-7} 
                                  &                                 & Siting     &     87.34$\pm$ 1.03      &      93.45$\pm$ 0.77       &      91.02$\pm$ 0.89       &    90.29$\pm$ 0.92      \\ \cline{3-7} 
                                  &                                 & Lying Down &     98.55$\pm$ 0.39      &      96.00$\pm$ 0.64       &      99.14$\pm$ 0.30       &    97.26$\pm$ 0.54      \\ \cline{2-7} 
                                  & \multirow{3}{*}{Non-Stationary} & Running    &     87.32$\pm$ 1.10      &      82.67$\pm$ 6.06       &      99.80$\pm$ 0.19       &    84.93$\pm$ 3.25      \\ \cline{3-7} 
                                  &                                 & Walking    &     73.97$\pm$ 1.44      &      69.12$\pm$ 7.39       &      98.30$\pm$ 1.01       &    71.46$\pm$ 4.10      \\ \cline{3-7} 
                                  &                                 & Bicycling  &     91.01$\pm$ 0.94      &      89.06$\pm$ 1.03       &      99.54$\pm$ 0.22       &    90.02$\pm$ 0.99      \\ \hline\hline
\multirow{6}{*}{\rotatebox[origin=c]{90}{HHAR-net}} & \multirow{3}{*}{Stationary}             & Standing   &     84.27$\pm$ 1.88      &      87.60$\pm$ 1.70       &      97.21$\pm$ 0.85       &                                       85.90$\pm$ 1.80     \\ \cline{3-7} 
                                  &                                 & Siting     &     91.42$\pm$ 0.87      &      96.02$\pm$ 0.61       &      94.15$\pm$ 0.73       &    93.66$\pm$ 0.76      \\ \cline{3-7} 
                                  &                                 & Lying Down &     98.84$\pm$ 0.35      &      98.49$\pm$ 0.40       &      99.30$\pm$ 0.27       &    98.66$\pm$ 0.37      \\ \cline{2-7} 
                                  & \multirow{3}{*}{Non-Stationary} & Running    &     89.76$\pm$ 0.99      &      84.44$\pm$ 6.11       &      99.86$\pm$ 0.13       &    87.02$\pm$ 3.12      \\ \cline{3-7} 
                                  &                                 & Walking    &     86.14$\pm$ 1.12      &      58.83$\pm$ 8.30       &      99.35$\pm$ 0.63       &    69.92$\pm$ 4.26      \\ \cline{3-7} 
                                  &                                 & Bicycling  &     92.96$\pm$ 0.83      &      86.07$\pm$ 1.13       &      99.68$\pm$ 0.18       &    89.38$\pm$ 1.00      \\ \hline
\end{tabular}
\vspace{12pt}
\caption{Detailed performance comparison of flat DNN and HHAR-Net. HHAR-Net exhibits a dominant performance over flat DNN}
\label{table:DetailedPerformaceComparison}
\end{table}


We trained a hierarchical DNN with the same hyper-parametrs that were used for the flat DNN. Then we tested our model on the test data as explained in the evaluation section. The resulting confusion matrix is shown in Figure \ref{fig:H_Conf_Mat}. Likewise, we calculated the misclassifications and summarized it into Table \ref{table:Hierarchical_Misclassification}. By comparing Tables \ref{table:Flat_Misclassifications} and \ref{table:Hierarchical_Misclassification}, we can see that the total misclassifications dropped significantly. Not only the number of misclassifications between stationary activities and non-stationary activities dropped from 393 to 338 but also the number of misclassifications within stationary classes decreased. This could be because in this hierarchical model, we are dealing with three-class classifiers at the parent level whereas the flat classifier has six classes.

Tables \ref{table:Accuracy_Benchmark} and \ref{table:Balanced_Accuracy_Benchmark} show the accuracy and balanced accuracy comparison of our proposed model, hierarchical DNN, and the baselines respectively. It is evident that our model is performing better than our baselines and also flat DNN. We achieved the accuracy of 95.8\% at the \(0^th\) level, differentiating between stationary and non-stationary activities. Moreover, the classification accuracy of the stationary and non-stationary activities at the parent level were 92.8\% and 93.2\% respectively resulting in a 92.8\% total accuracy.

To further verify the performance of HHAR-Net, we measured precision, sensitivity, specificity and F1 score for both flat DNN and HHAR-Net in table \ref{table:DetailedPerformaceComparison}. All of these metrics can be easily extracted from the confusion matrix nonetheless provided here for a more detailed comparison. One can witness that HHAR-Net shows better precision for all classes and also an improved Sensitivity for all classes except walking and bicycling. The same trend is seen in F1 score as it is a harmonic mean of precision and sensitivity. 
Another observation is that introducing a hierarchy has led to a significant improvement within the stationary classes which is our abundant class. On the other hand, non-stationary activities suffer from a lack of sufficient training data leading to larger error bars. This makes it hard to draw any conclusion about the slight superiority of the flat DNN F1 score over the HHAR-Net. Nevertheless, we can see a dominant performance of the proposed HHAR-Net over the flat DNN.

\section{Conclusion}

In this paper, we proposed a hierarchical classification for an HAR system, HHAR-Net. Our model proved to be capable of differentiating six activity labels successfully with a high accuracy surpassing flat classifiers, even a flat DNN with the same architecture. The idea of recognizing a structure in the labels can be applied in many other similar systems to improve the performance of the model and also make it more interpretable. 
Boosting human activity recognition facilitates the health monitoring of individuals in many scenarios. Based on a user’s activities and routines we can extract useful information about the user’s behavioral patterns and subsequently design and implement intervention systems to play a role in case of an abnormality.

Our approach can be extended by taking into account other possible structures in the context labels. Also, as the number of labels increases, discovering a structure and breaking down the classification task into smaller problems becomes more prominent and may call for efficient hierarchical methods.

%
%
%
\bibliographystyle{splncs04}
\bibliography{ref.bib}
\end{document}